\documentclass[preprint,showpacs,preprintnumbers,amsmath,amssymb]{revtex4}
\usepackage{graphicx}% Include figure files
\usepackage{dcolumn}% Align table columns on decimal point
\usepackage{bm}% bold math

\begin{document}

%\preprint{APS/123-QED}

\title{On the Dynamical Overstability of Radiative Blast Waves:\\
The Atomic Physics of Shock Stability}

\author{J. Martin Laming}
\email{jlaming@ssd5.nrl.navy.mil}
\affiliation{Space Science Division Code 7674L, Naval Research
Laboratory, Washington DC 20375}

\author{Jacob Grun}
\email{grun@nrl.navy.mil}
\affiliation{Plasma Physics Division
Code 6795, Naval Research Laboratory, Washington DC 20375}

\date{\today}% It is always \today, today,

\begin{abstract}
Atomic physics calculations of radiative cooling are used to
calculate criteria for the overstability of radiating shocks. Our
calculations explain the measurement of shock overstability by
Grun et al. and explain why the overstability was not observed in
other experiments. The methodology described here can be
especially useful in astrophysical situations where the relevant
properties leading to an overstability can be measured
spectroscopically, but the effective adiabatic index is harder to
determine.
\end{abstract}

\maketitle

Shocks play a crucial role in the death and rebirth of stars. At
the endpoint of stellar evolution, a supernova explosion launches
a blast wave out into the surrounding medium with a velocity in
the range 10,000 - 20,000 km s$^{-1}$. After about 10,000 years
the blast wave slows to speeds of order 200 km sec$^{-1}$ and
becomes ``radiative'', i.e. radiative energy losses from a
``cooling zone'' some distance behind the shock front itself
become an important consideration in the overall shock dynamics.
Radiative shocks are subject to a number of interesting
hydrodynamic instabilities and oscillations. A velocity dependent
cooling instability may develop as the shock slows
\cite{chevalier82,innes87,gaetz88,kimoto97}. This causes large
amplitude fluctuations in the shock velocity and in the distance
between the shock front and the radiative cooling zone. When the
shock slows sufficiently that the cooling instability dies away,
and the distance between the shock front and the cooling zone is
much smaller, it can become subject to a new instability, an
oscillatory rippling of its front that grows as a power of time
\cite{vishniac83,bertschinger86,ryu87,vishniac89,maclow93}. The
ripples grow because the thermal pressure of the shocked gas,
which is perpendicular to the local shock front, is not
necessarily parallel to the ram pressure of the upstream plasma,
which is directed along the shock velocity vector. In shocks with
sufficiently high compression this imbalance of pressures induces
oscillatory movement of material within the shock shell. Parts of
the shell that contain less mass slow down more than the parts of
the shell that contain more mass and a growing oscillation ensues.
In its nonlinear phase \cite{maclow93} knots or clumps of material
may form with sizes similar to the shocked shell thickness. It is possible
that local nonuniformities in interstellar gas caused by the
aforementioned instabilities provide the initial conditions for
gravitational collapse and the subsequent birth of new stars.

The existence of growing ripples in radiative shock
fronts was demonstrated in a laboratory experiment by Grun et al. \cite{grun91}.
These authors produced blast waves in nitrogen and xenon gas and
showed that shocks in the more radiative xenon gas rippled with a
power-law growth rate close to theoretical predictions, whereas
shocks in nitrogen remained stable. More recently other
researchers, working in a somewhat different parameter space,
attempted to produce the rippling overstability, but were unable to
do so \cite{edwards01}.

In this paper we perform detailed calculations of the radiative cooling of a shock
front in nitrogen and xenon plasma from which we derive the effective
adiabatic index $\gamma _{\rm eff}$
and using \cite{vishniac89} infer the growth rate of the
overstability, with the aim of understanding just how radiative a radiative
shock needs to be to be overstable.
We find that in practice in the laboratory examples we consider, the
shocked plasma must cool significantly
during the shock transition, which is a distance of order the ion mean free path
\cite{landau59,liberman86}. We compare our results to experiments
\cite{grun91,edwards01} and show why \cite{grun91} were able to observe the
overstability and why \cite{edwards01} could not. Since determining
$\gamma _{\rm eff}$ from
cooling calculations is dependent on quantities such as element abundances,
densities, temperatures and shock velocities that can be measured spectroscopically,
the formalism we present can be helpful in astrophysical situations where
the effective adiabatic index is otherwise much harder to determine.

Our method for computing the radiative cooling follows that in \cite{laming01}.
The Coulomb logarithm is set to its appropriate value, and three
body recombination is included, as is appropriate for high density plasma.
For N, we use collisional ionization and radiative and
dielectronic recombination rates from \cite{mazzotta98}. Ionization cross sections
for Xe$^{2-6+}$ and Xe$^{8+}$ can be found in
\cite{griffin84,gregory83a,gregory83b,bannister88,man93}, from which rates were
calculated and fitted to a Lotz formula. Rates for Xe, Xe$^+$ and Xe$^{7+}$ were
estimated by interpolation and extrapolation. We are unaware of
recombination rates for these Xe ions, so for radiative and
dielectronic recombination we substituted the corresponding rates
for Ar from \cite{mazzotta98}. This should not lead to large
errors, since the ground configurations of Xe and Ar are
5s$^2$5p$^6$ and 3s$^2$3p$^6$ respectively, and in most of the blast waves
we model, Xe does not ionize into the 4d$^{10}$ subshell. In any case three body
recombination is usually dominant. Radiative
cooling rates for N II and N III are calculated from collisional data given in
\cite{stafford94a,stafford94b}
and radiative data from \cite{galavis97,mendoza99,nussbaumer79} and the Opacity
Project. Cooling rates for more highly charged ions of N and for
the Xe ions were computed using the HULLAC suite of codes
\cite{barshalom88a,barshalom88b}. All cooling rates were tabulated
at a nominal electron density of $n_e=10^{17}$ cm$^{-3}$, and fitted
to formulae of the form $\left(1 + a\left(n_e/10^{17} -1\right)\right)^{-1}$ where
$a$ is a constant for each ion to model the density dependence. We neglect the
small temperature
dependence of $a$. The rates for N and Xe at $n_e=10^{17}$ cm$^{-3}$ were
similar to those tabulated in \cite{summers78} for N (interpolated
from those for C and O) and for Ar. The cooling rate for Xe X, the most highly
charged Xe ion in our model is the same as Ar X in \cite{summers78} multiplied
by a factor of 10, obtained by comparison with results in \cite{post77}.

Another modification made to \cite{laming01} is the use of a more
realistic density profile for the expanding laser target, though
this makes little difference to the blast wave evolution in the
Sedov-Taylor phase. With the appropriate ambient gas density and a
nominal ablated target kinetic energy of 100 or 200 J for N or Xe respectively
(coming from the
energy of the laser pulse \cite{grun91} ) we calculate the blast wave velocity
and radius as a function of time after the laser pulse. At each
time we compute a steady state radiative shock structure,
demanding that the photoionizing radiation produced by the shocked
gas must produce a self-consistent preshock ionization state.
Photoionization rates are taken from \cite{verner96}, with those
for Ar substituting for Xe. For neutral Xe at least, the
photoionization cross section is very similar to that for neutral
Ar \cite{henke88,chantler95}. We assumed the radiative cooling in
neutral N or Xe was ineffective due to opacity, and the
temperature of the preshock gas (the shock precursor) is
calculated by balancing the heating by photoionization with
radiative cooling. We also checked that molecular N$_2$ was
completely dissociated in the precursor by the radiation field
using photoionization/dissociation cross sections compiled at {\tt
http://www.space.swri.edu/amop/}, dissociative recombination
\cite{kella96}, electron impact dissociation
\cite{ajello96,stibbe99} and recombination \cite{clyne67}. In any
case our precursor temperatures are generally sufficiently high
that N$_2$ dissociation should not be an issue.

We model the shock interior by setting electron and ion temperatures equal
to the values given by the jump conditions added to their preshock
temperatures. We then follow a Lagrangian plasma element through the shock
by integrating the simultaneous equations for the
ionization balance and electron and ion temperatures accounting for
electron-ion collisional equilibration, radiation and ionization
energy losses. After each time step we modify particle
temperatures and number densities according to the effects of adiabatic
expansion of the blast wave, and radiative and ionization losses in an
assumed constant pressure environment. We proceed in this manner for a time 400 ns
following the laser pulse. Once the Lagrangian element has moved a distance $d$,
the shock width, given by \cite{landau59,liberman86}
$d=\left<\left(4/3\right)2\gamma /\left(\gamma +1\right)v_i\tau
_{ii}\right>$ we evaluate $\gamma $ at each time step from the density
enhancement of the plasma element relative to the preshock value.
In the expression for $d$, $v_i$ and $\tau _{ii}$ are the ion thermal
speed and self collision time respectively and the angled brackets $\left< ... \right>$
denote a time average through the shock transition.

\begin{figure}
\includegraphics{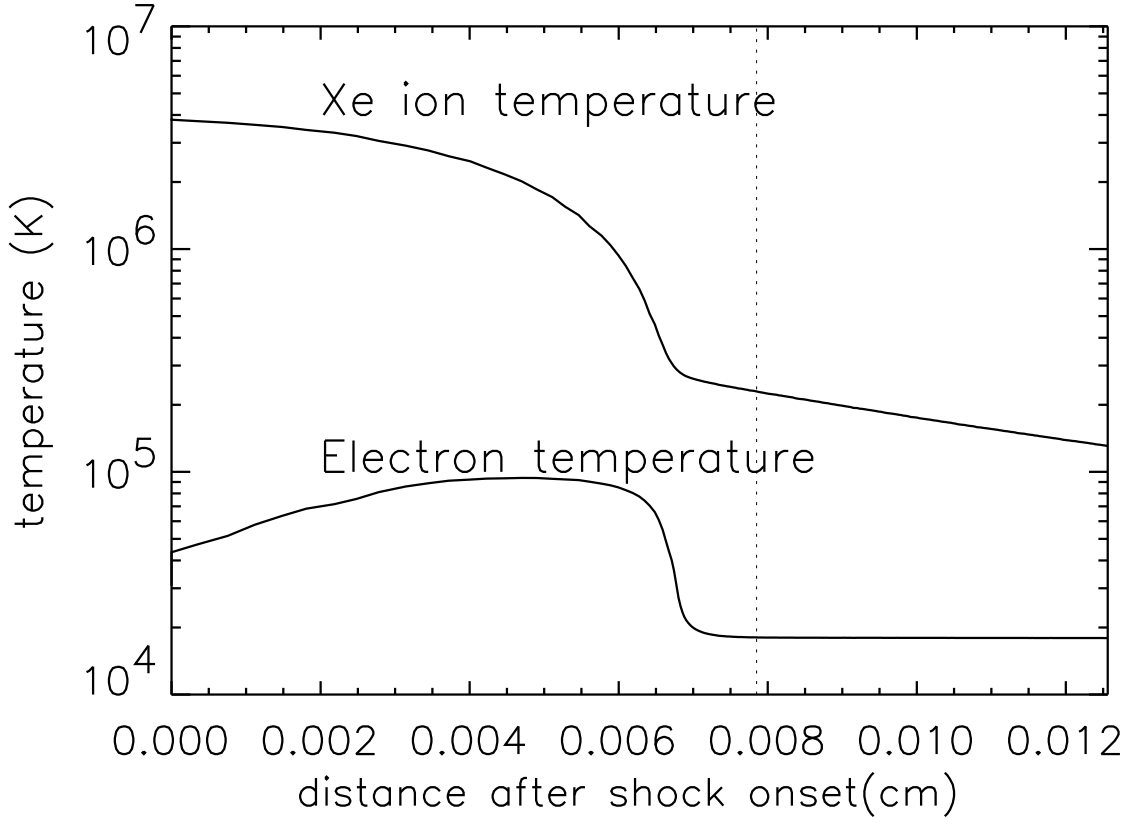}% Here is how to import EPS art
\caption{\label{temps} The spatial evolution of the electron and Xe ion temperatures
behind the shock front at 120 ns after laser pulse. Distance is measured from the
shock onset. The width of the shock is 7.85e-3 cm,
indicated by the vertical dotted line. }
\end{figure}

\begin{figure}
\includegraphics{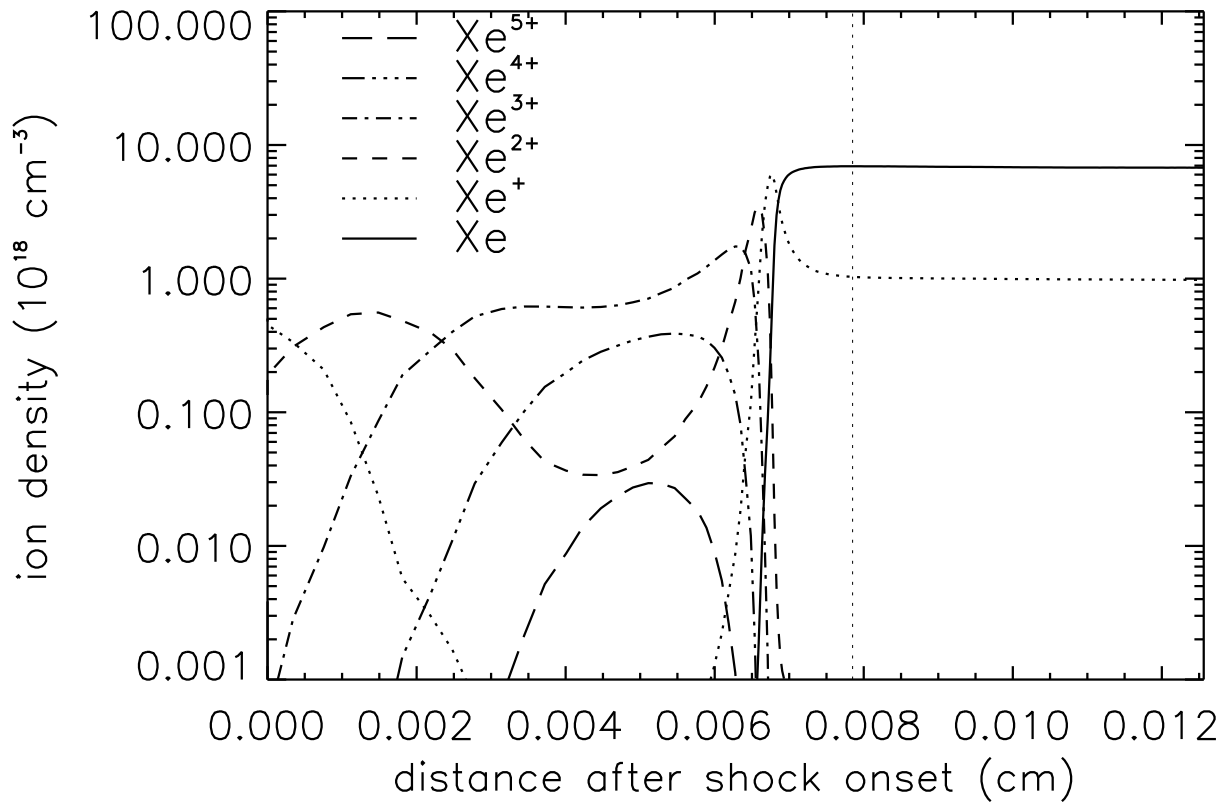}% Here is how to import EPS art
\caption{\label{ions} The spatial evolution of the Xe ionization balance at
120 ns after the laser pulse.}
\end{figure}

The evolution of the electron and ion
temperatures with distance behind the shock onset are plotted in Figure \ref{temps}
for the Xe blast wave 120 ns after the laser pulse, and the corresponding evolution
of the ionization balance is given in Figure \ref{ions}. We evaluate an average
$\gamma _{\rm eff}$ for various times in the evolution of the blast wave from
the average density enhancement in the accumulated shell of shocked gas over the
preshock density. These values are given in Tables
\ref{table1} for N and \ref{table2} for Xe. We also give at each time $t$ the shock
velocity $v_s$, radius $R$, the initial ionization state and temperature $T$,
and the Mach number $M$ appropriate to the precursor temperature $T$.


\begin{table*}
\caption{\label{table1} Model parameters for N blast wave, 100 J laser pulse.}
\begin{ruledtabular}
\begin{tabular}{cccccccccccc}
$t$ & $v_s$  & $R$ & N & N$^{+}$& N$^{2+}$& $T$ & $M$
 & $\gamma _{\rm eff}$ & real($s_{\rm max}$)& $l_{\rm max}$& $\epsilon$\\
(ns) & (km s$^{-1}$) & (cm) & & & & (K) &\\ \hline
 40& 87.5 & 0.68& 0.109& 0.845& 0.046& 32200& 15.6&  1.045 & -0.016$\to $0.32& 82$\to $114& 0.89\\
 80& 58.7 & 0.96& 0.211& 0.776& 0.013& 30100& 10.8&  1.20& -0.25$\to $ -0.065& & 0.49\\
120& 44.1 & 1.16& 0.377& 0.621& 0.002& 28100& 8.3&  1.26& & & 0.23\\
160& 36.3 & 1.32& 0.538& 0.462& 0.001& 26200& 7.1&  1.23& & & 0.17\\
200& 31.4 & 1.45& 0.684& 0.316& 0.0& 24800& 6.3&  1.22& & & 0.14\\
\end{tabular}
\end{ruledtabular}
\end{table*}

\begin{table*}
\caption{\label{table2} Model parameters for Xe blast wave, 200 J laser pulse.}
\begin{ruledtabular}
\begin{tabular}{cccccccccccccc}
$t$ & $v_s$  & $R$ &Xe & Xe$^{+}$& Xe$^{2+}$&
Xe$^{3+}$ & Xe$^{4+}$ &$T$ & $M$ & $\gamma _{\rm eff}$ & real($s_{\rm max}$)&
$l_{\rm max}$ & $\epsilon $\\
(ns) & (km s$^{-1}$) & (cm) & & & & & & (K) &\\ \hline
40& 75.3 & 0.60& 0.0& 0.040& 0.652& 0.281& 0.027& 46100& 34.2&  1.19& & &0.596\\
60& 57.3 & 0.73& 0.0& 0.263& 0.718& 0.019& 0.0 & 37400& 28.9& 1.015&
  0.60$\to $0.86& 257$\to $349& 0.963 \\
80& 47.0 & 0.84& 0.001& 0.439& 0.554& 0.006& 0.0 & 35100& 24.5& 1.025&
  0.13$\to $0.45& 152$\to $210& 0.942\\
100& 40.5& 0.93& 0.003& 0.647& 0.349& 0.001& 0.0 & 32600& 21.9& 1.028&
  0.13$\to $0.45& 135$\to $187& 0.918\\
120& 36.0 & 1.00& 0.006& 0.784& 0.210& 0.0 & 0.0 & 30700& 20.0& 1.033&
  0.046$\to $0.38 & 114$\to $158& 0.901\\
160& 29.9 & 1.13& 0.019& 0.917& 0.065& 0.0 & 0.0 & 27700& 17.5& 1.048&
  -0.17$\to $0.19& 77$\to $108& 0.876\\
200& 26.0 & 1.24& 0.047& 0.935& 0.018& 0.0 & 0.0& 25100& 16.0& 1.057&
  -0.23$\to $ 0.13& 64$\to $91 & 0.851\\
240& 23.2 & 1.34& 0.101& 0.894& 0.005& 0.0& 0.0 & 23100& 14.9& 1.075&
  -0.253$\to $0.021& $<69$ & 0.825\\
300& 20.2 & 1.47& 0.207& 0.792& 0.001& 0.0& 0.0 & 19000& 13.5&  1.096&
  $<-0.044$& & 0.775\\
400& 16.9 & 1.66& 0.406& 0.594& 0.0 & 0.0& 0.0 & 19000& 12.0&  1.13& $<-0.067$& &
0.710\\
\end{tabular}
\end{ruledtabular}
\end{table*}

\begin{figure}
\includegraphics{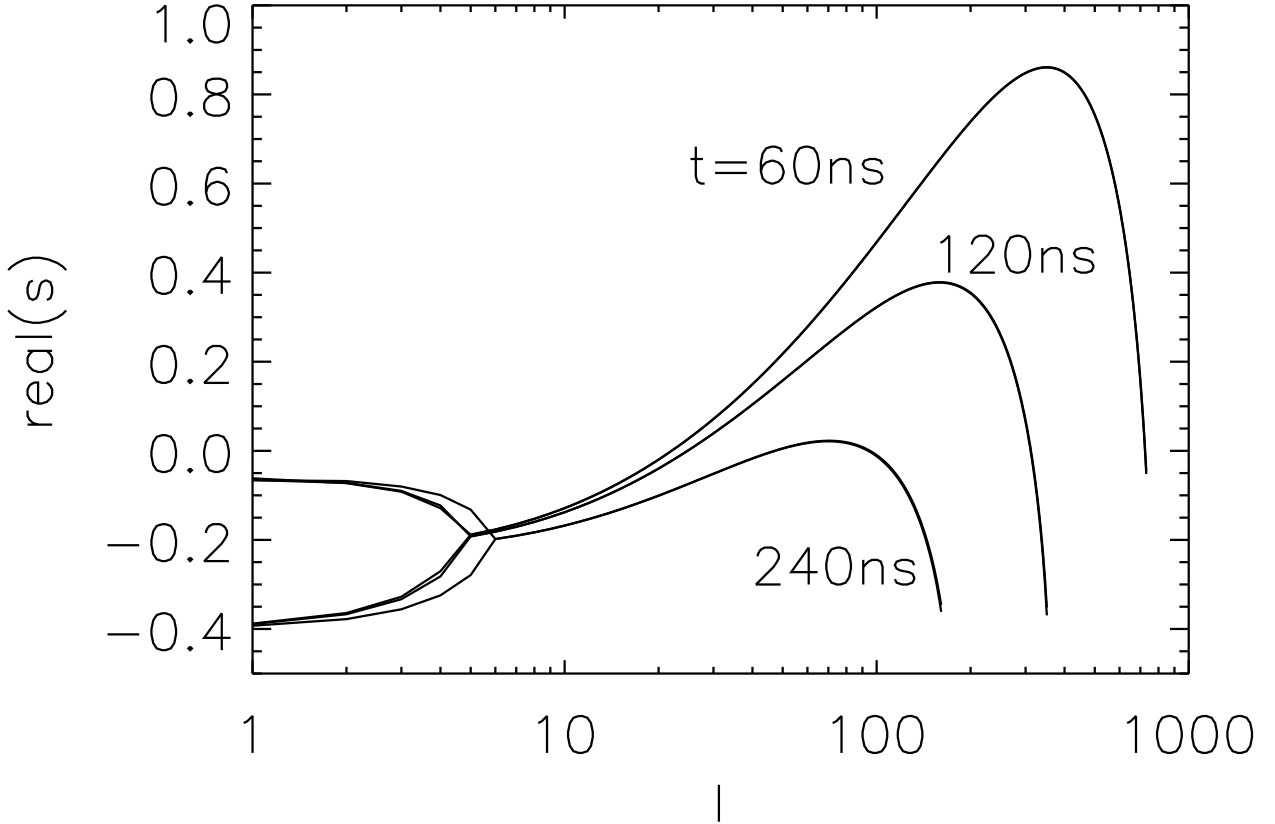}% Here is how to import EPS art
\caption{\label{vishniac} Plots of real($s$) where the overstability grows as
$t^s$, for Xe blast waves observed at 60, 120, and 240 ns after the laser pulse.
Curves are calculated from equations 19 of reference [6],%\cite[vishniac89],
using shock parameters given in Table \ref{table2}. }
\end{figure}

We evaluate the growth exponent real($s$) and the value of $l=kR$
for each blast wave from equations 19 in \cite{vishniac89}. The
maximum real($s$) and the $l$ at which this occurs
are given in the penultimate two columns of Tables \ref{table1}
and \ref{table2}. In this calculation we the shell thickness as a fraction of
the blast wave radius is taken to be
$H/R=\left(\gamma _{\rm eff}-1\right)/\left(\gamma _{\rm eff}+1\right)/3$.
The ranges of real($s$) and $l$ given in the tables correspond to taking
$R\propto t^m$ with $m=2/5$ for adiabatic Sedov-Taylor behavior or $m=2/7$
for the strongly radiating pressure driven snowplow case, which gives the
higher values of real($s$) and $l$. This is expected to be the case for
the blast wave under consideration, although the data of \cite{grun91}
appear to be slightly more consistent with $m=2/5$. However if the energy
radiated by the shocked plasma is absorbed upstream and
consequently swept back up by the shock, \cite{edwards01} speculate
that behavior closer to the
Sedov-Taylor limit may be observed even for strongly radiating
shocks, and this limit was assumed in the calculations of ionization balance
and radiative cooling.
Examples of the stability calculations are given in Figure \ref{vishniac} for
Xe at 60 ns, 120 ns, and 240 ns, which follow the transition
from strong overstability through to stability for $m=2/7$. We find generally
that $\gamma _{\rm eff}$ must be closer to 1 for overstability than in the
original work \cite{vishniac83}. This is because we calculate the Mach number
independently of $\gamma _{\rm eff}$ whereas \cite{vishniac83} couple them
to ensure an isothermal shock, as in equation (22) of \cite{vishniac89}. The
final column in Tables \ref{table1} and \ref{table2} gives the
fraction of the kinetic energy of the incident upstream plasma (in
the shock rest frame) that is radiated away during the shock
transition, $\epsilon$. This is estimated by identifying the $\gamma $ we
calculate at distance $d$ behind the shock with the $\gamma _1$ parameter in
\cite{cohen98,liang00}, and using their relation between $\epsilon$ and $\gamma _1$.

From Tables \ref{table1} and \ref{table2}, our results are in qualitative
agreement with the observations in \cite{grun91}. The N blast wave is
stable at all times (except at 40 ns, but here the blast wave evolution is probably
still dominated by the exploding target)
whereas the Xe blast wave shows overstability for times
between 60 and 150-300 ns, depending on the value of $m$.
The predicted stabilization for Xe at 150 - 300 ns
is in good agreement with observations. However we still
predict growth at lower $s$ and higher $l$ than actually observed, though
\cite{grun91} caution that their measurements of $kR$ may not be identical
to the $l$ in the theory. Additionally the theory only treats the linear
regime, while the measurements presumably include non-linear
effects.

The overstability is suppressed at early times because insufficient time has elapsed
to allow the shock heated Xe plasma to cool significantly.
Similar speed blast waves launched by a more energetic laser pulse (and thus
having decelerated from a higher initial velocity with more time available
for cooling) would be overstable. At late times the blast wave
stabilizes simply because the radiative power loss becomes insufficient at the
lower shock speeds. From Table
\ref{table2} it appears that approximately 80-90\% of the incident plasma kinetic energy
must be radiated in the shock transition before overstability occurs.
It is also now clear why overstabilities were not
observed in \cite{edwards01}. They launched shocks in Xe gas at atmospheric
pressure with velocities initially 15 km s$^{-1}$ slowing to around 6 km
s$^{-1}$. The blast waves in \cite{grun91} were initiated at
speeds of order 100 km s$^{-1}$ in Xe gas at 5 torr pressure. From
Table \ref{table2} it is clear that at 5 torr pressure below a minimum shock speed of
around 25 km s$^{-1}$ Xe will no longer be overstable. This
minimum shock speed will likely be higher for Xe shocks at
atmospheric pressure, since the higher density will reduce
radiative cooling rate by electron collisional depopulation of
excited levels.

We believe that we have captured the essential physics of the radiative blast
waves observed in \cite{grun91}. A more rigorous treatment must dispense with the fluid
approximation and use a kinetic theory description of the plasma.
However such a calculation with the necessary atomic physics is
probably some years away in terms of the computing resources required.
The fundamental reason why the N blast waves are
stable is not so much that N is inherently less radiative
than Xe at the relevant temperatures, but that its radiation is
more suppressed in our cooling calculations by the electron
density than that for Xe. However
we do still expect that heavy element plasmas, rather than the H-He
dominated cosmic composition, will be more susceptible to the
overstability. Thus a promising astrophysical environment in which
to look for such effects might be the heavy element rich plasma in
the ejecta of supernova remnants, for which the reverse shock can
be radiative in early phases \cite{chevalier94}.

\begin{acknowledgments}
We acknowledge support by basic research funds of the Office of
Naval Research (JML), as well as support from NRL 6.2 Award
N0001402WX30007
and ONR Award N0001402WX20803 (JG). We thank Dr Judah Goldwasser
from ONR for his support an encouragement.
\end{acknowledgments}

%\newpage %Just because of unusual number of tables stacked at end
\bibliography{grun91}% Produces the bibliography via BibTeX.

\end{document}